\documentclass[hyper,letterpaper,12pt]{article}
\usepackage{a4wide}
\usepackage{amsmath}
\usepackage[
      colorlinks=true,
      linkcolor=blue,
      urlcolor=blue,
      filecolor=black,
      citecolor=blue,
      pdfstartview=FitV,
        pdfpagemode=None,
        bookmarksopen=true
      ]{hyperref}
\usepackage{color}
\usepackage{graphicx}
\usepackage{amsfonts}
\usepackage{amssymb}

\def\bc{\begin{center}}
\def\nno{\nonumber}
\def\ec{\end{center}}
\def\be{\begin{eqnarray}}
\def\ee{\end{eqnarray}}

\definecolor{dyellow}{rgb}{1.,0.8,.0}
\definecolor{myblue}{rgb}{.1,.1,.7}
\definecolor{dcyan}{rgb}{.0,.6,.6}
\definecolor{dmagenta}{rgb}{0.6,0.0,0.6}
\definecolor{brown}{rgb}{0.6,0.2,0.}
\definecolor{darkblue}{rgb}{.0,.0,0.5}
\definecolor{darkred}{rgb}{0.75,0.0,0.0}
\definecolor{orange}{rgb}{1.,.6,.0}
\definecolor{dorange}{rgb}{0.8,.4,.0}
\definecolor{darkgreen}{rgb}{0.0,0.6,0.0}
\definecolor{purple}{rgb}{.4,.0,.4}
\definecolor{lightgrey}{rgb}{0.7, 0.7, 0.7}
\definecolor{grey}{rgb}{0.4, 0.4, 0.4}
\def\al{\alpha}

\def\eps{\epsilon}

\def\la{\lambda}

\def\om{\omega}

\def\pa{\partial}

\def\La{\Lambda}

\def\r{\partial}

\begin{document}

\title{\bf Phase transitions in AdS soliton spacetime through marginally stable modes}

\author{
~Rong-Gen Cai\footnote{E-mail: cairg@itp.ac.cn}~$^{{ 1}}$, ~Xi
He\footnote{E-mail: 20100083@hznu.edu.cn}~$^ 2$,~ Huai-Fan
Li\footnote{E-mail: huaifan.li@stu.xjtu.edu.cn}~$^{ 1,3,4}$,
~Hai-Qing Zhang\footnote{E-mail: hqzhang@itp.ac.cn}~$^{{ 1}}$
\\
\\
$^{{1}}$\small Key Laboratory of Frontiers in Theoretical Physics,\\
\small Institute of Theoretical Physics, Chinese Academy of Sciences,\\
\small P.O. Box 2735, Beijing 100190, China\\
$^{2}$\small Department of Physics, Hangzhou Normal University, \\
\small Hangzhou 310036,
China\\
 $^{3}$\small
Department of Physics and Institute of Theoretical
   Physics,\\
\small Shanxi Datong University, Datong 037009, China \\
 $^{4}$\small Department of
Applied Physics, Xi' an Jiaotong University, \\
\small Xi' an 710049, China}

\date{\small (\today)}

\maketitle

~~~~~~~~~{\footnotesize PACS numbers: 11.25.Tq, 74.20.-z, 04.70.-s}

\begin{abstract}
\normalsize

We investigate the marginally stable modes of the scalar (vector)
perturbations in the AdS soliton background coupled to electric
field. In the probe limit, we find that the marginally stable modes
can reveal the onset of the phase transitions of this model. The
critical chemical potentials obtained from this approach are in good
agreement with the previous numerical or analytical results.
\end{abstract}

\section{ Introduction}

The AdS/CFT correspondence~\cite{Maldacena:1997re} provides a
powerful theoretical method to understand the strongly coupled field
theories. Recently it has been applied to study the holographic
models of superconductors (superfluids) phase transitions since the
work~\cite{Gubser:2008px,Hartnoll:2008vx}. For reviews see
\cite{Hartnoll:2009sz}. Since the condensed matter physics deals
with the systems at finite charge and finite temperature, from the
AdS/CFT correspondence the dual gravity should be described by a
charged black hole.

The phase transition studied in~\cite{Gubser:2008px,Hartnoll:2008vx}
is actually a holographic superconductor/metal phase transition. The
 model for holographic superconductors can be simply constructed by
an Einstein-Maxwell theory coupled to a complex scalar field. In
particular, when the temperature of the black hole is below a
critical temperature, the black hole solution becomes unstable to
develop a scalar hair near the horizon. And the condensation of the
scalar hair breaks the U(1) symmetry of the system. From the AdS/CFT
correspondence, the complex scalar field is dual to a charged
operator in the boundary field theory. And the breaking of the U(1)
symmetry in gravity causes a global U(1) symmetry breaking in the
dual boundary theory. This induces a superconductor (superfluid)
 phase transition \cite{Weinberg:1986cq}.

The holographic insulator/superconductor phase transition was first
studied in \cite{Nishioka:2009zj}. In particular, they used a
five-dimensional AdS soliton background \cite{Horowitz:1998ha}
coupled to a Maxwell and scalar field to model the holographic
insulator/superconductor phase transition at zero temperature. The
normal phase in the AdS soliton is dual to a confined gauge theory
with a mass gap which resembles an insulator phase
\cite{Witten:1998zw}. When the chemical potential is sufficiently
large beyond a critical value, the AdS soliton  becomes unstable to
form scalar hair which is dual to a superconducting phase in the
boundary field theory. The holographic insulator/superconductor
phase transition was also studied in
\cite{Horowitz:2010jq,Akhavan:2010bf,Basu:2011yg,Brihaye:2011vk,
Cai:2011ky}.

To reveal the stability of a spacetime background, a powerful method
is to study the quasinormal modes (QNMs) of the perturbations in
this background, for reviews see \cite{Kokkotas:1999bd,
Berti:2009kk,Konoplya:2011qq}. If the imaginary part of the QNMs is
negative, the mode will decrease in time and the perturbation will
finally disappear which indicates that the background is stable
against this perturbation. On the contrary, if the imaginary part of
the QNMs is positive, this implies that the background is unstable
against this perturbation. The interesting thing is that if the
perturbation has a marginally stable mode, {\it i.e.} $\om=0$, one
always expects that this is a signal of instability, or rather a
phase transition may occur. For the detailed discussions of
marginally stable modes, one can refer to \cite{Gubser:2008px}.

In this paper, we employ the idea of marginally stable modes to
study the s-wave and p-wave holographic insulator/superconductor
phase transitions in AdS soliton background at zero temperature. The
construction of the system is like the one in
\cite{Nishioka:2009zj}. We take advantage of Horowitz and Hubeny's
method \cite{Horowitz:1999jd} to study the QNMs of the scalar
(vector) perturbations of this system. By increasing the chemical
potential from zero to some critical value, the marginally stable
modes will turn out. This means at the critical chemical potential
the AdS soliton background becomes unstable and will prefer to be an
AdS soliton background coupled with nonzero charged scalar (vector)
fields. This argument is consistent with the previous studies on
holographic insulator/superconductor phase transitions
\cite{Nishioka:2009zj,Horowitz:2010jq, Akhavan:2010bf}. In
particular, the critical chemical potentials we get from the
marginally stable modes are in good agreement with the ones in
\cite{Nishioka:2009zj,Horowitz:2010jq, Akhavan:2010bf}. Actually,
there are multiple marginally stable modes corresponding to various
critical chemical potentials. These modes are related to the modes
of node $n=1, 2, 3 \cdots$. However, they are unstable due to the
oscillations of scalar (vector) field in the radial direction
\cite{Gubser:2008px, Gubser:2008wv}. By making use of the
alternative approach, {\it viz.} ``shooting" method, we plot the
behavior of the scalar (vector) fields depending on the radial
direction. From these diagrams, one can intuitively see the ``nodes"
of these fields. The studies of QNMs in AdS soliton background are
also investigated in \cite{Shen:2007xk,He:2010zb}.

The paper is organized as follows. In Sect.\eqref{sect:swave}, we
study the marginally stable modes of s-wave field in the probe
limit. We find that the critical chemical potentials we derived are
consistent with the results obtained by previous works. By making
use of the same procedure, we explore the QNMs of p-wave field in
the AdS soliton background in Sect.\eqref{sect:pwave}. We draw our
conclusions and make some discussions in Sect.\eqref{sect:con}.
\section{S-wave perturbations}
\label{sect:swave}

Following Ref.\cite{Nishioka:2009zj}, we set up this model in the
AdS soliton background \cite{Horowitz:1998ha}:
 \be \label{metric} ds^2=L^2\frac{dr^2}{f(r)}+r^2(-dt^2+dz^2+dy^2)+f(r)d\chi^2.\ee
where, $f(r)=r^2-r_0^4/r^2$ and $L$ is the radius of AdS spacetime.
In fact, this soliton solution can be obtained from a
five-dimensional Schwarzschild-AdS  black hole by making use of two
Wick rotations. The asymptotical AdS space-time approaches to a
topology of $R^{1,2}\times S^1$  near the boundary. And the
Scherk-Schwarz circle $\chi\sim\chi+\pi L/r_0$ is required in order
to have a smooth geometry. The geometry looks like a cigar whose tip
is at $r=r_0$. Because of the compactified direction $\chi$, this
background provides a gravity description of a three-dimensional
field theory with a mass gap, which resembles an insulator in the
condensed matter physics. The temperature in this background is
zero.

It is well known that in this AdS soliton background a simple
solution for Maxwell gauge field is $A_t={\rm Const.}=\mu$. Instead
of $A_t=0$ at the horizon required by the AdS black holes, $A_t$ can
be any non-singular value at the tip of the AdS soliton.

In the probe limit, we introduce a charged scalar field $\psi$ as a
probe into this background which is a neutral AdS soliton with a
constant electric potential. The Lagrangian for the charged scalar
field is
 \be \mathcal{L}_{\rm matter}=-|\nabla_{\mu}\psi-iqA_{\mu}\psi|^2-m^2|\psi|^2.\ee
The Euler-Lagrange equations of motion (EoMs) for $\psi$ is
 \be\label{eompsi} (\nabla_\mu-iqA_\mu)(\nabla^\mu-iqA^\mu)\psi-m^2\psi=0.\ee
In the following, we  assume that $\psi$ is real and
$\psi=F(t,r)H(\chi)Y(z,y)$. Substituting $\psi$ into
Eq.\eqref{eompsi} and making the separation of the variables we can
reach
 \be\label{Feom} &&\frac{\pa^2F(t,r)}{\pa r^2}+(\frac3r+\frac{\pa_rf}{f})\frac{\pa F(t,r)}{\pa
 r}-\frac{L^2}{fr^2}\frac{\pa^2F(t,r)}{\pa t^2}+\frac{2iq\mu
 L^2}{fr^2}\frac{\pa F(t,r)}{\pa
 t}\nno\\&&+\frac{L^2}{fr^2}(q^2\mu^2-m^2r^2-\frac{\la^2r^2}{f}-\xi^2)F(t,r)=0.\ee
where $\la$ and $\xi$ are the eigenvalues of the following equations
respectively
 \be \frac{\pa^2 H(\chi)}{\pa\chi^2}+\la^2H(\chi)&=&0,\\
     \frac{\pa^2Y(z,y)}{\pa z^2}+\frac{\pa^2Y(z,y)}{\pa
     y^2}+\xi^2Y(z,y)&=&0.\ee
where $\la=2r_0n/L$, $n\in \mathbb{Z}$ due to the periodicity of
$H(\chi)=H(\chi+\pi L/r_0)$ and $\xi\in \mathbb{Z}$. For simplicity
we will set $\la=\xi\equiv0$ which means that there are no momenta
in the $(z,y,\chi)-$ directions.

\subsection{Critical behavior from marginally stable modes}

Further, we define $F(t,r)=e^{-i\om
 t}R(r)$ and Eq.\eqref{Feom} becomes (we have set $L\equiv1$)
 \be\label{Reom}
 R~''(r)+(\frac{f'}{f}+\frac3r)R~'(r)+\frac{1}{fr^2}[(\om+q\mu)^2-m^2r^2]R(r)=0.\ee
where a prime denotes the derivative with respect to $r$.

In order to investigate the phase transitions of this model, we
recall that the marginally stable modes can to some extent reveal
this critical behavior \cite{Gubser:2008px}. In the studies of QNMs,
marginally stable modes correspond to $\om=0$ which indicates that
the phase transition or the critical phenomena may occur. We will
take advantage of the Horowitz and Hubeny's method
\cite{Horowitz:1999jd} to study these QNMs.

It is convenient to convert the $r$-coordinate to $x$-coordinate,
where $x=1/r$. Therefore, the infinite boundary is now at $x=0$
while the tip is at $x=x_0=1/r_0$. In terms of this new coordinate
$x$, Eq.(\ref{Reom}) becomes
 \be x^4 \pa_{xx}{R(x)}+\bigg[-x^3+\frac{x^4\pa_x
 f(x)}{f(x)}\bigg]\pa_xR(x)+\frac{1}{f(x)}\bigg[x^2(\om+q\mu)^2-m^2\bigg]R(x)=0.\ee
Following the steps of Horowitz and Hubeny \cite{Horowitz:1999jd},
we can multiply $f(x)/(x-x_0)$ to both sides of the above equation,
and we reach
 \be \label{maineq}
 S(x)
 \pa^2_xR(x)+\frac{T(x)}{x-x_0}\pa_xR(x)+\frac{U(x)}{(x-x_0)^2}R(x)=0,\ee
 where the coefficient functions are given by
 \be S(x)&=&\frac{f(x)x^4}{x-x_0},\\
 T(x)&=&-f x^3+x^4\pa_x
 f(x),\\
 \label{potential}U(x)&=&\bigg[x^2(\om+q\mu)^2-m^2\bigg](x-x_0).\ee
Note that $x=x_0$ is a regular singular point of $S(x), T(x)$ and
$U(x)$, and we can polynomially expand them to a finite order like
 \be\label{sexpand} S(x)=\sum_{n=0}^M s_n(x-x_0)^n.\ee
where $M$ is a finite integer. The series expansion of $T(x)$ and
$U(x)$ can be similarly reduced.

Unlike the ingoing boundary conditions of scalar field near a black
hole horizon, the boundary conditions here can be a finite quantity
at the tip of the AdS soliton. We can expand $R(x)=(x-x_0)^\al$ and
substitute it into Eq.\eqref{maineq}.  Then to the leading order we
get
 \be \al(\al-1)s_0+\al t_0+u_0=-4x_0\al^2=0 \Longrightarrow \al=0.\ee
This corresponds to looking for a solution of the form
 \be\label{expand} R(x)=\lim_{N\rightarrow\infty}\sum_{n=0}^{N}a_n(x-x_0)^n.\ee
Substituting \eqref{expand} and \eqref{sexpand} into \eqref{maineq}
and comparing the coefficients of $(x-x_0)^n$ for the same $n$ we
find that
 \be\label{an} a_n&=&-\frac{1}{P_n}\sum_{k=0}^{n-1}[k(k-1)s_{n-k}+k
 t_{n-k}+u_{n-k}]a_k,\\
 \label{pn} P_n&=&n(n-1)s_0+nt_0+u_0=-4x_0n^2.\ee
We set $a_0=1$ due to the linearity of Eq.\eqref{maineq}. The
boundary conditions for the scalar field at $x=0$ is
  \be\label{R0} R(0)=\lim_{N\rightarrow\infty}\sum_{n=0}^{N}a_n(-x_0)^n=0.\ee
And the algebraic equation \eqref{R0} can solve the modes $\om$.

In the following numerical calculations, we restrict $x_0=1$ and
$q=1$ just like the ones in \cite{Nishioka:2009zj}. In practice, we
will expand $R(x)$ to a large order which is $N=300$. In order to
find the marginally stable modes $\om=0$ of the system, we should
restrict our attention to the potential $U(x)$ in \eqref{potential}.
One finds that $\om$ and $\mu$ are symmetric ($q=1$). This means
that when $\mu=0$ the lowest-lying modes of $\om$ will exactly be
identical to the lowest-lying critical chemical potentials $\mu_c$
when $\om=0$.\footnote{Here, we only take care of the QNMs with
 positive real part. Therefore, the lowest-lying modes of $\om$
 represent the modes which have the minimal or less minimal positive real parts. } Using this trick, we
can easily find the critical chemical potentials where the
marginally stable modes arise.

\begin{figure}[h]
 \centering
\includegraphics[scale=0.9]{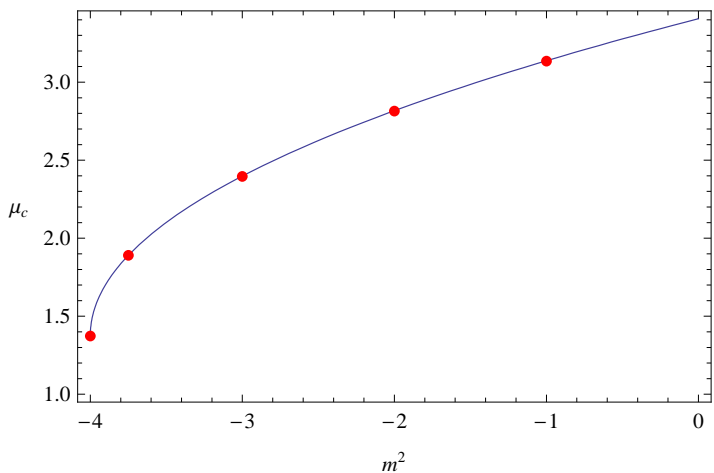}
\caption{\label{mum2} The critical chemical potentials for the
marginally stable modes versus $m^2$ of the scalar field. The blue
curve is taken from the analytical results in \cite{Cai:2011ky}
while the red points are numerically obtained from the QNMs. They
are perfectly matched.}
\end{figure}

 Fig.\eqref{mum2} shows the critical chemical potentials $\mu_c$'s for
 various squared mass of the scalar field. The red points are obtained from the
 numerical calculations of the marginally stable modes while the
 blue curve is taken from the analytical results we have previously
 derived in \cite{Cai:2011ky}. They agree with each other perfectly. In particular,
when $m^2=-15/4$, $\mu_c\approx 1.885$ obtained from the marginally
stable modes is in good agreement with the result $\mu_2=1.88$ which
is the critical value for the onset of the holographic
insulator/superconductor phase transitions obtained in
\cite{Nishioka:2009zj}. This implies that the marginally stable
modes can indeed reveal the
 phase transitions of this model.

\begin{table}[h]
\caption{\label{tab} The first three lowest-lying critical chemical
potentials $\mu_c$ for various mass squares obtained from the
calculation of the marginally stable modes. The overtone numbers of
them are $n= 0, 1, 2$ from the top to the bottom.}
\begin{center}
\begin{tabular}{cccc}
  \hline
 $~~m^2=-4~~$ & $~~m^2=-15/4~~$ & $~~m^2=-3~~$ & $~~m^2=-2~~$ \\
  \hline
 $1.373$ & $1.885$ & $2.396$ & $2.815$ \\
  $3.658$ & $4.226$ & $4.792$ & $5.246$ \\
 $6.029$ & $6.603$ & $7.189$ & $7.655$ \\
  \hline
\end{tabular}
\end{center}
\end{table}

Table \eqref{tab} shows the first three lowest-lying critical
chemical potentials for various $m^2$'s. The $\mu_c$'s of the
overtone number $n=0$ are the critical chemical potentials shown in
Fig.\eqref{mum2}.  Other critical chemical potentials of overtone
numbers $n=1, 2$ can also make the QNMs to be marginally stable.
However, they are expected to be unstable which can be understood
after the next subsection. The nodes $n=0, 1, 2$ can also be
intuitively seen in the next subsection.

\subsection{Critical behavior from the ``shooting" method}

Actually there is another way to study the critical behavior of this
phase transition. It was called the ``shooting" method which was
commonly used in the previous studies on holographic superconductors
\cite{Hartnoll:2009sz}. Here, we will concisely describe how to make
use of this ``shooting" method to study the critical behavior in AdS
soliton background and  compare it with the results of ``marginally
stable modes" method.

We will work in the approximation that $A_t={\rm Const.}=\mu$ and
the scalar field only depends on $r$-direction as well as that it is
too small to back-react the background. The EoMs of $\psi(r)$ is
 \be\label{psieomr}
 \psi''(r)+(\frac{f'}{f}+\frac3r)\psi'(r)+(\frac{\mu^2}{r^2f}-\frac{m^2}{f})\psi(r)=0.\ee
The boundary condition of $\psi(r)$ at the tip is
 \be \psi=a + b (r-r_0)+\cdots,\ee
and near the infinite boundary $\psi(r)$ behaves as
 \be \psi=\frac{\psi^{(1)}}{r^{2-\sqrt{4+m^2}}}+\frac{\psi^{(2)}}{r^{2+\sqrt{4+m^2}}}+\cdots\ee
In the following calculations, we will set $\psi^{(1)}=0$ in order
to turn off the effect of the source on the boundary field theory.
\footnote{It is well known that when $0<\sqrt{4+m^2}<1$ the scalar
admits two different quantizations~\cite{Klebanov:1999tb}. In this
case, $\psi^{(1)}$ can either be a source or an expectation value
according to the standard quantization or the alternative
quantization, respectively. In our paper, we will only focus on the
standard quantization.}

The ``shooting" method states that we can start with a initial value
of $\psi$ at the tip $r_0$ and then perform the numerical
calculations of the EoMs of $\psi$ Eq.\eqref{psieomr} provided that
the infinite boundary conditions $\psi^{(1)}=0$ are satisfied. At
the critical point of the phase transition, the quantity of $\psi$
is very close to zero. Therefore, we have set the initial value of
$\psi$ to be $0.01$ in our numerical calculations.

 \begin{figure}[h]
 \centering
\includegraphics[scale=0.7]{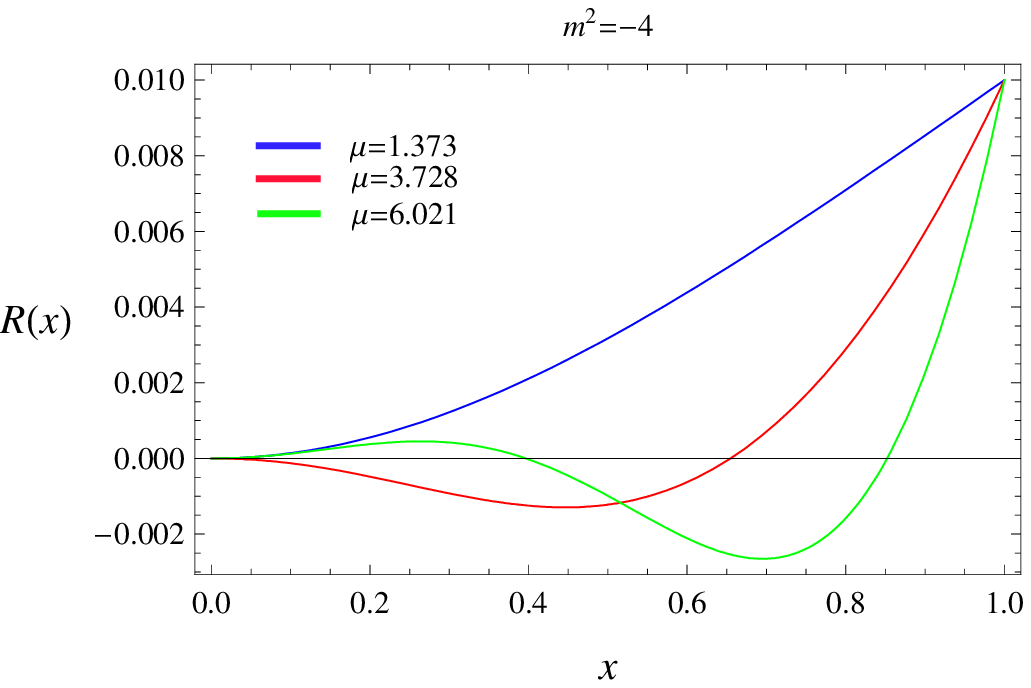}
\includegraphics[scale=0.7]{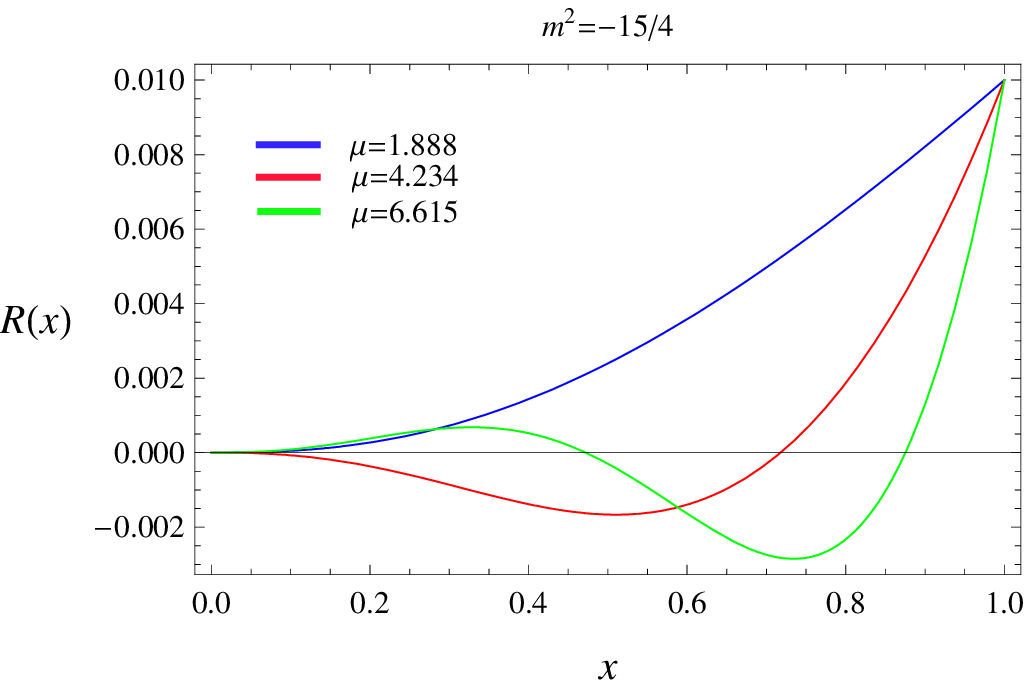}\\
\includegraphics[scale=0.7]{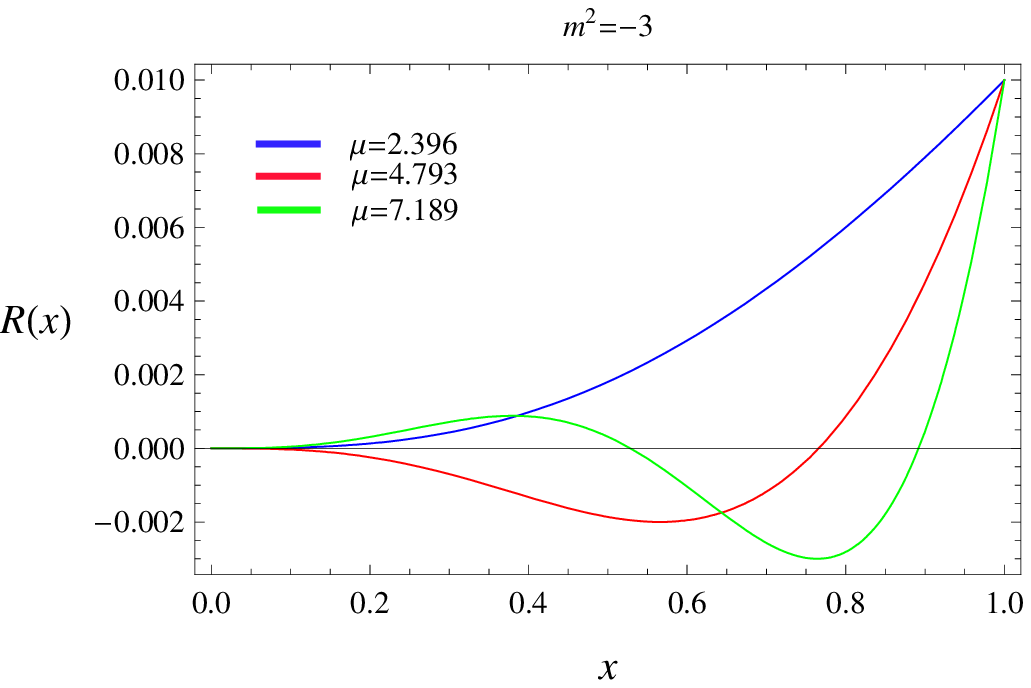}
\includegraphics[scale=0.7]{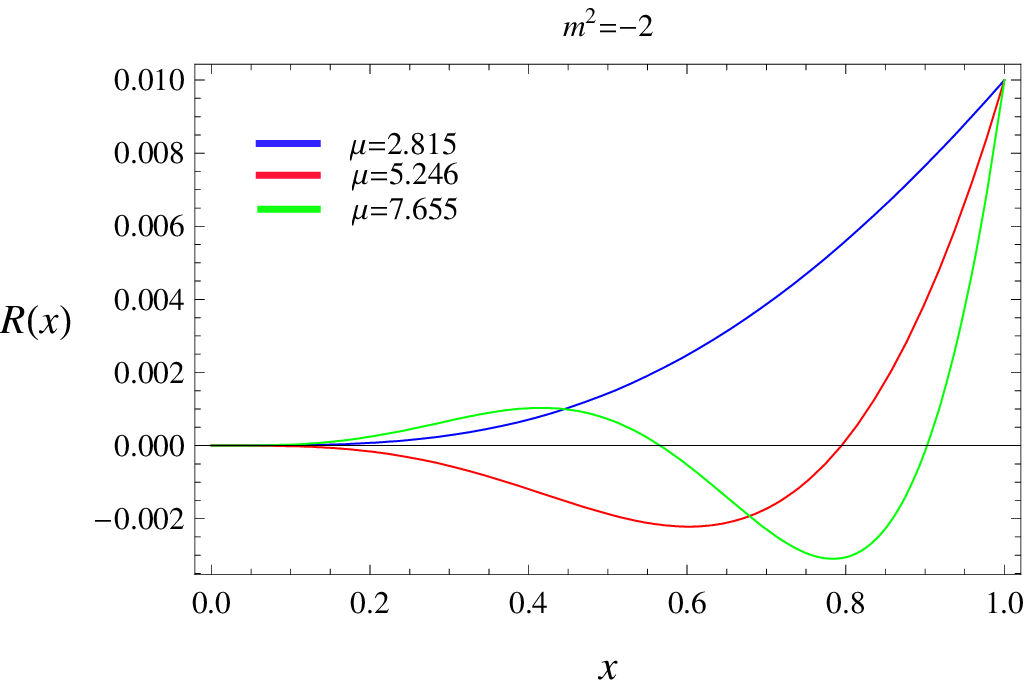}
\caption{\label{nodes} The marginally stable curves of scalar fields
corresponding to various critical chemical potentials in the cases
of different squared mass. The critical chemical potentials for
different curves are in the sequence $\mu_c^{\rm blue}<\mu_c^{\rm
red}<\mu_c^{\rm green}$.}
\end{figure}

Fig.\eqref{nodes} shows the multiple marginally stable curves of the
scalar fields $R(x)$\footnote{Note that we have transformed the
scalar field $\psi(r)$ to $R(x)$ in $x=1/r$ coordinates.} in $x=1/r$
coordinates for various $m^2$. Take the plot of $m^2=-15/4$ for
example, the first three lowest-lying chemical potentials are in the
sequence $\mu_c^{\rm blue}<\mu_c^{\rm red}<\mu_c^{\rm green}$. We
find that the values of $\mu_c$'s obtained from the ``shooting"
method are perfectly consistent with the $\mu_c$'s in Table
\eqref{tab} derived from the ``marginally stable modes" method. The
blue line is for the minimal value of $\mu_c$. It starts from a very
small initial value at the tip $x=1$ and then monotonically decrease
to zero at the infinite boundary $x=0$. There are no other
intersecting points between the blue line and the $R(x)=0$ axis.
This is the reason we call the mode with $\mu_c\approx1.888$ of node
$n=0$. For the red and green lines, they are related to
$\mu_c=4.234$ and $\mu_c=6.615$ respectively. The red line has one
intersecting point with $R(x)=0$ axis while the green line has two.
Therefore, as we have mentioned above, we can call the two modes
with nodes $n=1$ and $n=2$ respectively. However, the red and green
lines are expected to be unstable, because radial oscillations in
$x$-direction of $R(x)$ will cost energy \cite{Gubser:2008wv}. In
addition, the arguments above are also appropriate for other
diagrams in Fig.\eqref{nodes}.

\section{P-wave perturbations}
\label{sect:pwave}

In this section we extend our work to study an AdS soliton with
p-wave vector fields. We consider a five-dimensional SU(2)
Einstein-Yang-Mills theory with a negative cosmological constant
following \cite{Gubser:2008zu} . The action is
 \be S=\int d^5x
 \sqrt{-g}[\frac12(R-\La)-\frac14F^a_{\mu\nu}F^{a~\mu\nu}],\ee
where $F^a_{\mu\nu}$ is the field strength of the SU(2) gauge theory
and
$F^a_{\mu\nu}=\r_{\mu}A^a_{\nu}-\r_{\nu}A^a_{\mu}+\eps^{abc}A^b_{\mu}A^c_{\nu}$.
$a, b, c=(1,2,3)$ are the indices of the SU(2) Lie algebra
generator. $A^a_{\mu}$ are the components of the mixed-valued gauge
fields $A=A^a_{\mu}\tau^adx^{\mu}$, where $\tau^a$ are the
generators of the SU(2) Lie algebra with commutation relation
$[\tau^a, \tau^b]=\eps^{abc}\tau^c$. And $\eps^{abc}$ is a totally
antisymmetric tensor with $\eps^{123}=+1$.

As a consistent solution of the system, in the probe limit, the
background of the metric can also be an AdS soliton solution like
\eqref{metric} (We have scaled $L\equiv1, r_0\equiv1$),
 \be
ds^2= \frac{dr^2}{r^2g(r)}+r^2(-dt^2+dz^2+dy^2)+r^2g(r)d\chi^2,
\label{eq1} \ee where we have set $f(r)=r^2g(r)=r^2(1-1/r^4)$.

We adopt the ansatz for the gauge field as~\cite{Gubser:2008wv}
  \be\label{ansatz} A(t,r)=\phi(r)\tau^3dt+\psi(t,r)\tau^1dz.\ee
Note that in order to consider the QNMs of the vector field, we have
assumed $\psi(t,r)$ depends on $t$ and $r$. In this ansatz, the
gauge boson with nonzero component $\psi(t,r)$ along $z$-direction
is charged under $A^3_t=\phi(r)$. According to AdS/CFT dictionary,
$\phi(r)$ is dual to the chemical potential in the boundary field
theory while $\psi(t, r)$ is dual to the $z$-component of some
charged vector operator $\hat O$. The condensation of $\psi(t, r)$
will spontaneously break the U(1)$_3$ gauge symmetry and induce the
phenomena of superconducting on the boundary field theory.

Define $x=1/r$, the EoMs for $\phi(x)$ and $\psi(t,x)=e^{-i\om
t}R(x)$ in the $x$ coordinate are
 \be
 \label{phieomp}\phi''+(\frac{g'}{g}-\frac1x)\phi'-\frac{R^2}{g}\phi=0,\\
 \label{Reomp} R~''+(\frac{g'}{g}-\frac1x)R~'+\frac{\om^2+\phi^2}{g}R=0.\ee
where a prime denotes the derivative with respect to $x$. A simple
solution is that $\phi(x)={\rm Const.}=\mu$ and $R(x)=0$ which
corresponds to an neutral AdS soliton with a constant electric
potential.

In order to find the marginally stable modes of $R(x)$, we can
follow the steps in the previous section. Eq.\eqref{Reomp} can be
transformed to
 \be\label{mainp}
 S(x)\pa_x^2R(x)+\frac{T(x)}{x-1}\pa_xR(x)+\frac{U(x)}{(x-1)^2}R(x)=0,\ee
where, the coefficient functions are
 \be S(x)&=&\frac{gx^4}{x-1},\\
     T(x)&=&x^3(-g+x \pa_xg),\\
    \label{potentialp} U(x)&=&x^4(\om^2+\mu^2)(x-1).\ee
Polynomially expand these three coefficient functions to a finite
order and $R(x)$ to an infinite order, such as
 \be \label{sp} S(x)&=&\sum_{n=0}^{M}s_n(x-1)^n,\\
     \label{rp} R(x)&=&\lim_{N\rightarrow\infty}\sum_{n=0}^{N}a_n(x-1)^n.\ee
where $M$ is a finite integer. Substitute \eqref{rp} and \eqref{sp}
 into Eq.\eqref{mainp} we reach a recursion relation as
Eqs.\eqref{an} and \eqref{pn} in the previous section. The solutions
of $\om$ can be derived by requiring $R(0)=0$.

In the practical numerical calculations, we take the order of the
expansion of $R(x)$ to be $N=300$. Notice again that in the
potential $U(x)$ \eqref{potentialp} $\om$ and $\mu$ are symmetric.
Using the trick we have adopted in the preceding section, we can
easily find the marginally stable modes $\om=0$ when $\mu=\mu_c$.
The first three lowest-lying critical chemical potentials are
 \be \mu_c\approx2.265, 4.742, 7.156. \ee
 The minimal critical chemical
 potential $\mu_c\approx2.265$ is in perfect agreement with the
 results in \cite{Akhavan:2010bf,Cai:2011ky}.

In order to study the behavior of the field $R(x)$, we should first
know the boundary conditions for $R(x)$ and then make use of the
``shooting" method to numerically calculate it. Notice again that we
still assume the electric field $A_t={\rm Const.}=\mu$ here. At the
tip, $R(x)$ behaves as
 \be R(x)=a+b(x-1)+\cdots,\ee
while
 \be R(x)=R^{(0)}+R^{(1)} x^2+\cdots.\ee
at the infinite boundary. In the following calculation we will set
$R^{(0)}=0$ in order not to source the field theory on the boundary.

\begin{figure}[h]
 \centering
\includegraphics[scale=0.9]{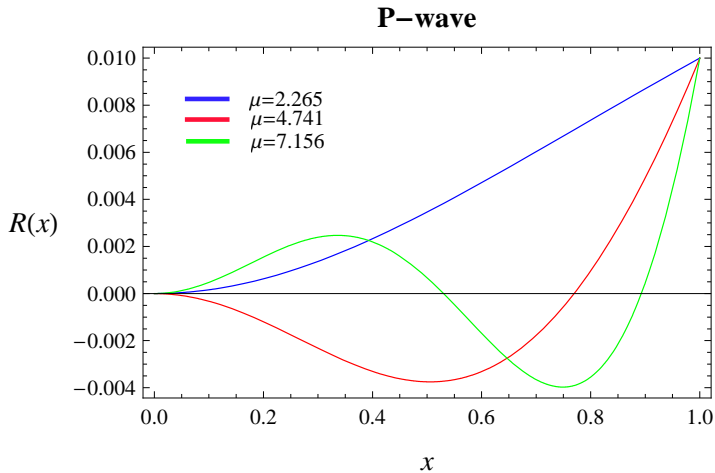}
\caption{\label{pwavenode} The marginally stable curves of vector
fields $R(x)$ corresponding to various chemical potentials. The
critical chemical potentials for different curves are in the
sequence $\mu_c^{\rm blue}<\mu_c^{\rm red}<\mu_c^{\rm green}$.}
\end{figure}

We can see from Fig.\eqref{pwavenode} that there are multiple
marginally stable curves corresponding to different critical
chemical potentials. The blue line is related to the minimal value
$\mu_c=2.265$ which is obtained from the ``shooting" method. Once
again, we find that this critical value of chemical potential is
identical to the one derived from the ``marginally stable modes"
method. In addition, other values of $\mu_c$  are also consistent.
From what we have learned in the last section, the blue line
corresponds to the modes of node $n=0$ while red and green lines are
related to the modes of node $n=1$ and $n=2$ respectively. However,
the last two modes, {\it i.e.}, $n=1$ and $n=2$ are unstable due to
the cost of energy.

\section{Discussions and Conclusions}
\label{sect:con}

In this paper, we have studied the marginally stable modes for the
s-wave and p-wave perturbations in AdS soliton background with a
constant electric potential. At some critical chemical potentials,
the marginally stable modes, {\it i.e.} $\om=0$ will arise. The
importance of the marginally stable modes is that they can reveal
the instabilities of the background which in our paper means that
the neutral AdS soliton  will become unstable to develop charged
scalar (vector) ``hairs" in this AdS soliton background. Although
the detailed phase transitions cannot be seen in the study of QNMs,
it actually has been announced in the previous works
\cite{Nishioka:2009zj,Horowitz:2010jq, Akhavan:2010bf} by the study
of thermodynamics such as the free energy. This phase transition in
the gravity side will map to an insulator/superconductor phase
transition on the boundary field theory.

Despite that we do not exactly know the phase structures through the
marginally stable modes, they can actually indicate the onset of the
phase transition. This has been argued by Gubser in studying the
holographic superconductor phase transitions
\cite{Gubser:2008px,Gubser:2008wv}. In particular, marginally stable
modes can be obtained by studying the QNMs of the perturbations. The
widely used method to study QNMs in asymptotically AdS spacetime was
Horowitz and Hubeny's method \cite{Horowitz:1999jd}. In this paper,
we have adopted this method to find  that at some critical chemical
potentials, there indeed appeared the marginally stable modes. These
critical chemical potentials were in good agreement with those
obtained from the previous studies \cite{Nishioka:2009zj,
Akhavan:2010bf, Cai:2011ky}. We also took advantage of the
``shooting" method to numerically plot the behaviors of the scalar
(vector) fields in the radial direction. Therefore, one can
intuitively see the ``nodes" of the marginally stable modes. In
addition to the modes of the minimal critical chemical potentials,
we also studied other less lowest-lying marginally stable modes by
both methods. We asserted that these less lowest-lying marginally
stable modes are unstable because the oscillations of the scalar
(vector) fields in radial direction will cost energy.

\section*{Acknowledgements}
We would like to thank Hong-Bo Zhang for his indispensable help. HFL
would be very grateful for the hospitalities of the members in the
Institute of Theoretical Physics, Chinese Academy of Sciences. This
work was supported in part by the National Natural Science
Foundation of China (No. 10821504, No. 10975168, No.11035008 and
No.11075098), and in part by the Ministry of Science and Technology
of China under Grant No. 2010CB833004.

%\appendix

\end{document}